%% file: sbom.tex
\documentclass[letterpaper,twocolumn,10pt]{article}
\usepackage{usenix2020}

\usepackage{graphicx,grffile} %
\usepackage{comment}
\usepackage{color}
\usepackage{soul}
\usepackage{fancyvrb}
\usepackage{array}
\usepackage{booktabs}
\usepackage{listings}
\usepackage{multirow}
\usepackage{subcaption}
\usepackage[most]{tcolorbox}
\usepackage[bottom,marginal,flushmargin,hang]{footmisc}

\definecolor{bblue}{RGB}{30, 90, 170}
\definecolor{softorange}{RGB}{230, 126, 34}
\definecolor{freshgreen}{RGB}{46, 160, 67}
\definecolor{lightgray}{gray}{0.8}%
\definecolor{aaltoBlack}{RGB}{0,0,0}%
\definecolor{aaltoGray}{RGB}{146,139,129}%
\definecolor{aaltoGrayScale}{gray}{0.45}%
\definecolor{aaltoRed}{RGB}{237,41,57}%
\definecolor{aaltoBlue}{RGB}{0,101,189}%
\definecolor{aaltoYellow}{RGB}{254,203,0}%
\definecolor{aaltoPurple}{RGB}{102,57,183}%
\definecolor{aaltoTurquoise}{RGB}{0,168,180}%
\definecolor{aaltoGreen}{RGB}{0,155,58}%
\definecolor{aaltoLightGreen}{RGB}{105,190,40}%
\definecolor{aaltoOrange}{RGB}{255,121,0}%
\definecolor{aaltoFuchsia}{RGB}{177,5,157}%

\usepackage[T1]{fontenc}
\PassOptionsToPackage{hyphens}{url}
\usepackage{hyperref}
\hypersetup{breaklinks=true}

\makeatletter 
\patchcmd\caption@subtypehook{\let\label\subcaption@label}%
{\let\label\subcaption@label\let\ltx@label\subcaption@label}{}{\fail}
\makeatother

\newcommand{\codefontsize}{\fontsize{8}{9.75}\selectfont}

\lstdefinestyle{dpkg}{
  aboveskip=0pt,
  belowskip=0pt,
  keywordstyle=\color{aaltoBlue},
  keywords={Package,Status,Architecture,Source,Version,Depends,Priority,Section,Maintainer},
  classoffset=1
}

\lstdefinestyle{apk}{
  aboveskip=0pt,
  belowskip=0pt,
  keywordstyle=\color{aaltoBlue},
  keywords={T,L,o,c,r,R,Z},
  classoffset=1
}

\makeatletter
\lst@InstallKeywords k{attributes}{attributestyle}\slshape{attributestyle}{}ld
\makeatother

\lstdefinestyle{spdx}{
    aboveskip=0pt,
    belowskip=0pt,
    escapeinside={(*@}{@*)},
    keywordstyle=\color{aaltoBlue},
    keywords={spdxVersion,dataLicense,name,creationInfo,created,packages,files,relationships},
    literate=
      {\"}{{{\color{aaltoGrayScale}{\textquotedbl}}}}{1}
      {\{}{{{\color{aaltoGrayScale}{\{}}}}{1}
      {\}}{{{\color{aaltoGrayScale}{\}}}}}{1}
      {[}{{{\color{aaltoGrayScale}{[}}}}{1}
      {]}{{{\color{aaltoGrayScale}{]}}}}{1},
    classoffset=1,
    attributestyle = \color{aaltoGreen}, %
}

\usepackage[scaled=.95]{inconsolata}
\usepackage{todonotes}
\setuptodonotes{size=\tiny,backgroundcolor=yellow,linecolor=gray,noline}
\tikzset{/tikz/notestyleraw/.append style={node font=\ttfamily}}

\newcommand{\note}[3][Anon.]{\hl{#2}\todo{\textbf{#1}: #3}}
\usepackage{marginnote}
\newcommand{\fnote}[3][Anon.]{\hl{#2}\marginnote{\todo[inline]{\textbf{#1}: #3}}}

\renewcommand{\note}[3]{}\renewcommand{\fnote}[3]{}

\usepackage{wasysym}
\newcommand{\na}{\textcolor{gray}{--}}%
\newcommand{\no}{$\times$}%
\newcommand{\partially}{\LEFTcircle}
\newcommand{\yes}{\CIRCLE}

\AtBeginDocument{\urlstyle{tt}}

\usepackage[outline]{contour}
\contourlength{0.3pt}
\usepackage{pdftexcmds}
\makeatletter
\newcommand*{\mystrcmp}[2]{%
  \expandafter\pdf@strcmp\expandafter{#1}{#2}%
}
\makeatletter
\newcommand{\plotfont}{\sffamily\fontsize{6}{6.75}\selectfont}
\newcommand{\xscaling}{.925}
\newcommand{\heatmap}[4][-30]{%
  \begin{tikzpicture}[xscale=0.5525,yscale=.4]
  \plotfont
  \newcommand{\heat}{}
  \foreach \y [count=\n] in {#4}{
    \foreach \x [count=\m] in \y {
      \ifnum\mystrcmp{\x}{-}=-0 %
        \renewcommand{\heat}{lightgray}
      \else
        \pgfmathtruncatemacro{\shade}{#1*ln(\x+0.00001)}
        \renewcommand{\heat}{yellow!\shade!aaltoGreen}
      \fi
      \node[fill=\heat,minimum height=4mm, minimum width=5.75mm, text=white] at (\m,-\n) {\scalebox{\xscaling}[1.0]{\contour{black}{\textbf{\x}}}};
    }
  }
  \foreach \a [count=\i] in {#3} {
      \node[minimum  width=8mm,rotate=30] at (\i-.55,-7.375) {\parbox{8mm}{\hfill\scalebox{0.975}[1.0]{\a}}};
  }
  \foreach \a [count=\i] in {#2} {
    \node[minimum  width=8mm] at (-0.375,-\i) {\parbox{8mm}{\hfill\scalebox{0.95}[1.0]{\a}}};
  }
  \end{tikzpicture}
}

\usepackage{pgfplots}

\begin{document}

\title{SBOMproof: Beyond Alleged SBOM Compliance for Supply Chain Security of Container Images}

\author{
{\rm Jacopo Bufalino}\\
Cnam, Cedric, Paris, France\\
Aalto University, Espoo, Finland\\
jacopo.bufalino@cnam.fr
\and
{\rm Mario Di Francesco}\\
Aalto University, Espoo, Finland\\
mario.di.francesco@aalto.fi
\and
{\rm Agathe Blaise}\\
Thales SIX GTS France,\\ Gennevilliers, France\\
agathe.blaise@thalesgroup.com
\and
{\rm Stefano Secci}\\
Cnam, Cedric, Paris, France\\
stefano.secci@cnam.fr
}

\newcounter{observationcounter}
\setcounter{observationcounter}{1} %
\newcommand{\observationfontsize}{\fontsize{8.75}{10}\selectfont}
\newcommand{\observationfont}{\observationfontsize\sffamily}
\newtcolorbox{observation}{
  colback=gray!10,    %
  colframe=gray!70,   %
  boxrule=0.5mm,      %
  arc=1mm,            %
  left=2mm,           %
  right=2mm,          %
  top=1mm,            %
  bottom=1mm,         %
  before upper={\refstepcounter{observationcounter}}, %
  title={Observation~\theobservationcounter},
  fontupper=\itshape\observationfont,
  fonttitle=\bfseries\observationfont,
  code={\linespread{1.025}}
}

\maketitle

\input{abstract}

\input{introduction}

\input{background}

\input{methodology}

\input{compliance}

\input{accuracy}

\input{vulnerability}

\input{discussion}

\input{related}

\input{conclusion}

\bibliographystyle{plain}
\bibliography{sbom.bib}

\appendix
\include{appendix}

\end{document}

%% file: abstract.tex
\begin{abstract}
Supply chain security is extremely important for modern applications running at scale in the cloud. In fact, they involve a large number of heterogeneous microservices that also include third-party software. As a result, security vulnerabilities are hard to identify and mitigate before they start being actively exploited by attackers. For this reason, governments have recently introduced cybersecurity regulations that require vendors to share a software bill of material (SBOM) with end users or regulators. An SBOM can be employed to identify the security vulnerabilities of a software component even without access to its source code, as long as it is accurate and interoperable across different tools. This work evaluates this issue through a comprehensive study of tools for SBOM generation and vulnerability scanning, including both open-source software and cloud services from major providers. We specifically target software containers and focus on operating system packages in Linux distributions that are widely used as base images due to their far-reaching security impact. Our findings show that the considered tools are largely incompatible, leading to inaccurate reporting and a large amount of undetected vulnerabilities. We uncover the SBOM confusion vulnerability, a byproduct of such fragmented ecosystem, where inconsistent formats prevent reliable vulnerability detection across tools.
\end{abstract}

%% file: introduction.tex
\section{Introduction}

Modern applications %
run at scale in the cloud as a large number of microservices managed by an orchestrator, in most cases Kubernetes\cite{Casalicchio:2019:SMET,kubernetes}. The software in a microservice is packaged as a container image, namely, a self-contained artifact that includes both an application and all the components it requires, including operating system (OS) libraries and binaries. Because of this, containers are also subject to sophisticated attacks such as the XZ Utils backdoor~\cite{OPENWALL:2024:ONLINE}, which injected obfuscated code into an SSH server to allow for unauthorized access and arbitrary code execution. The attack leveraged social engineering and specifically targeted that package because of its widespread adoption in the Linux ecosystem, with potentially catastrophic consequences~\cite{Akamai:xzutilsbackdoor:web}. %
Such a striking example demonstrates the importance of software supply chain security (SSC), namely, the process of managing and mitigating the risks associated with any party involved in the development, distribution, and maintenance of applications and digital services~\cite{Williams:2025:TESM}. In particular, it highlights how attacks are perpetrated through components that are not the core of the developed software, but rather ``happen'' to be jointly distributed with it.

But how to exactly know what software is included in an artifact? Answering this question entails to accurately identify all the components in a software artifact, including its dependencies such as open-source libraries and third-party products~\cite{COFANO:2024:ARXIV,DUAN:2021:NDSS,LADISA:2023:IEEESP}. Such a process is automated by tools that generate a Software Bills of Material (SBOM), namely, a machine-friendly representation of a software artifact~\cite{SPDX:ONLINE}. An SBOM can be used to identify and mitigate security vulnerabilities through specialized tools that query multiple databases (e.g., security trackers)~\cite{trivy,snyk,syft,AWS:Inspector:web,AzureSBOM,google_sbom}. This process is absolutely necessary when the source code is not available~\cite{Dalia:2024:ARES,BENEDETTI:2024:ARXIV}. For these reasons, recent mandates from different governments -- including the US executive order on cybersecurity~\cite{GOVUS:2024:NIST} and the EU Cyber Resilience Act~\cite{CRA:2024:EU} -- require companies to share an SBOM with the end users or the regulator~\cite{NTIA:2019:SBOM}. %
As a consequence, SBOM interoperability is a primary concern; however, practitioners have recognized data quality, exchange formats, and tooling as major challenges in the adoption and effective use of SBOMs in SSC security~\cite{Zahan:2023:IEEESP}. Unfortunately, existing research has marginally addressed these challenges (Section~\ref{sec:related}).

To fill this gap, this work specifically addresses the interoperability of different SBOM tools along with their impact on SSC security. Specifically, we provide a comprehensive study of open-source software and cloud services from major providers for SBOM generation and vulnerability scanning of software containers (Section~\ref{sec:dataset}). Motivated by attacks such as the XZ Utils backdoor, we specifically focus on OS packages by considering Debian and Alpine, the most widely used Linux distributions for containers. Our findings highlight fundamental differences in package and CVE identification across tools. Moreover, package identifiers in SBOMs are not interoperable, eventually leading to inaccurate vulnerability reports. Our study also uncovered a novel vulnerability, the SBOM confusion, where incompatibilities in the SBOM prevent reliable vulnerability detection across tools. We finally implement a translation layer %
that addresses such an issue and is effective in improving the interoperability of the SBOMs across different tools.

In detail, this work establishes the following contributions.
\begin{itemize}
    \item It thoroughly characterizes the compliance of widely used tools to generate SBOMs for software containers (Section~\ref{sec:compliance}). Our evaluation considers the conformance to the Package URL (pURL) specification~\cite{purl:spec} and the SPDX format~\cite{SPDX:ONLINE}, which are both prevalent in the SBOM tooling landscape.
    \item It analyzes the impact of the SBOM structure on the vulnerability reports in detail (Section~\ref{sec:accuracy}). Our findings show that the considered tools are largely incompatible, leading to inaccurate reporting and a large amount of undetected vulnerabilities.
    \item It introduces a the SBOM confusion vulnerability, a novel security issue resulting from a fragmented SBOM ecosystem, where inconsistent outputs prevent reliable vulnerability detection across tools (Section~\ref{sec:vuln}). We also develop \texttt{sbomvert}, an open-source software\footnote{\scalebox{.9375}[1.0]{The code is available at \url{https://anonymous.4open.science/r/sbomvert-759D}} and will be publicly released with the final version of this paper.} that translates a tool-specific SPDX file into one that can effectively be used for vulnerability scanning.
    \item A thorough discussion of the issues involved in the end-to-end process ranging from SBOM generation to vulnerability scanning (Section~\ref{sec:discussion}). We also consider the impact of CVE formats on security assessment and the emerging solution of zero-CVE container images.
\end{itemize}

%% file: background.tex
\section{Background}
\label{sec:background}

This section overviews the fundamental concepts in the context of software supply chain security, with a focus on those that are relevant for cloud environments. 

\begin{figure*}[tp!]
    \centering
    \subcaptionbox{\label{fig:dpkg}}[.4825\textwidth]{%
\begin{lstlisting}[style=dpkg]
Package: passwd
Status: install ok installed
Priority: required
Section: admin
Maintainer: Shadow package ...
Architecture: i386
Source: shadow (1:4.13+dfsg1-1)
Version: 1:4.13+dfsg1-1+b1
Depends: libaudit1 ...
\end{lstlisting}%
    }
    \quad
    \subcaptionbox{\label{fig:apk}}[.4825\textwidth]{%
\begin{lstlisting}[style=apk]
T:Alpine base dir structure and init scripts
L:GPL-2.0-only
o:alpine-baselayout
c:5f0cd7890349e7fe11128478ac506c709805224d
r:alpine-baselayout
...
R:passwd
Z:Q1r+bLonZkAyBix/HLgSeDsez22Zs=
\end{lstlisting}%
    }
    \caption{Fragments of a \subref{fig:dpkg} \texttt{dpkg} and an \subref{fig:apk} \texttt{apk} entry for the \texttt{passwd} package.}\label{fig:packages}
\end{figure*}

\subsection{Software Composition Analysis}

Software Composition Analysis (SCA) refers to the process of automatically identifying the packages, versions, licenses, and dependencies of a software artifact~\cite{SCA:2024:GITHUB}. The goal of SCA is to detect issues related to license compliance, outdated references, and security vulnerabilities. SCA is particularly useful for software that relies on third-party dependencies~\cite{COFANO:2024:ARXIV,BENEDETTI:2022:SCORED}, and tools are generally specialized for different ecosystems. One such case is represented by package management systems (equivalently, package managers), namely, tools that automate the installation, upgrade, and removal of software components. They maintain a local database of installed packages and their versions; they also interact with centralized repositories to determine if the installed packages are up-to-date, automatically updating those that are not. There are package managers for both programming languages (for instance, \texttt{npm} for JavaScript) and operating systems (typically Linux). Relying on package managers makes SCA more scalable as it can be statically performed by scanning metadata and index files instead of individual binaries (e.g., executables and libraries). This is especially important for complex software projects with a large number of components, which generally produce sizeable artifacts.

\subsection{Container Images and OS Packages}

Modern software runs in the cloud as a set of microservices managed through an orchestrator such as Kubernetes~\cite{RedHat:2024:web}. Each microservice is defined by a container image, which specifies the software as a self-contained artifact with all the needed dependencies. A container image comprises different layers that are applied on top of each other. The first one is a \emph{base image}, namely, a filesystem as the minimal building block required for the rest of the software in a microservice~\cite{ARXIV:2015:Bui}. Base images are typically represented by Linux operating system distributions such as Debian. The related software is generally installed through distribution-specific package managers. As a consequence, the image contains a possibly large amount of binaries in the base image~\cite{CLUSTER:2019:Zhao}. For these reasons, SCA for containers relies on the operating systems' package managers to efficiently identify packages and libraries. SCA for container images can either analyze individual layers of the image, or more often, the resulting filesystem (i.e., after combining all layers in the order they are specified).

The most widely employed images in cloud deployments are based on the Debian and Alpine distributions~\cite{DEVISSER:2017:ONLINE}, which use the \texttt{dpkg} and \texttt{apk} package managers, respectively (Figure~\ref{fig:packages}). The \texttt{dpkg} package manager stores state information in the \texttt{/var/lib/dpkg/status} file. Each installed package has an entry in this file, containing metadata such as the package name, version, architecture, and dependencies (Figure~\ref{fig:dpkg}). Each software component may be divided into a source package (or upstream) and one or more binary packages -- \texttt{dpkg} creates a separate entry in the status file for each binary package. For instance, Figure~\ref{fig:dpkg} shows that the binary package \texttt{passwd} was built from the source package \texttt{shadow}. Similarly, the \texttt{apk} package manager stores information about the installed packages in \texttt{/lib/apk/db/installed}. Also in this case, the package source is stored as part of the package information (Figure~\ref{fig:apk}).

\subsubsection{Package versioning and maintenance}

Debian and Alpine distribute third-party software using custom versioning schemas that may differ from the upstream project.
Debian provides three main release types: stable, testing, and unstable (\texttt{sid}). Stable releases have frozen package versions and receive only security updates and critical bug fixes.
Alpine maintains a rolling \texttt{edge} branch and periodically creates stable release branches from it. Stable branches receive only security updates and bug fixes. While Alpine generally preserves the upstream version numbering across branches, Debian appends release-specific revision numbers that may differ between releases (e.g., the \texttt{tiff} library has version \textit{4.2.0-1+deb11u5} in Debian 11 \texttt{bullseye} and \textit{4.5.0-6+deb12u2} in Debian 12 \texttt{bookworm}).

\subsection{Software Bill of Materials}

A Software Bill of Materials (SBOM) is an inventory of all components, libraries, licenses, and dependencies present in a software artifact. SBOMs are an important document for developers to track the health of their software, and also a legal requirement. Moreover, an accurate SBOM enables to identify and mitigate vulnerabilities associated with third-party components. There are well-established standards for SBOM generation, such as CycloneDX~\cite{CycloneDX:ONLINE} and SPDX~\cite{SPDX:ONLINE}. These standards define the format and structure of the SBOM, including information such as component names, versions, licenses, and relationships. This information helps understand the composition of the software and assess its security posture. In particular, each component or package has an identifier, namely, a unique string that embeds all the necessary information to pinpoint a certain artifact. Identifiers play a crucial role in SBOMs as they enable accurate mapping of software components to their corresponding vulnerabilities.

\subsubsection{The pURL package identifier}

Package URL (pURL) is an initiative led by the open-source community in collaboration with the OWASP Foundation. pURL identifiers are generated following a well-defined set of rules (see the example in Figure~\ref{fig:purl}) in a format that is currently being standardized as an ECMA specification~\cite{ecma:purl}.

\begin{figure}[b!]
    \centering
    \begin{lstlisting}
pkg:type/namespace/name@version?qualifiers#subpath
pkg:npm/angular/animation@12.0.0?platform=linux#bundles
    \end{lstlisting}
    \caption{pURL format (above) and an example pURL (below).}\label{fig:purl}
\end{figure}

The mandatory information for a pURL identifier includes: the \texttt{type}, the \texttt{name}, and the \texttt{version} of a package. The specification also optionally allows a \texttt{namespace} field, which represents either the author of the package (e.g., \texttt{angular} or \texttt{react}) or a subdomain of the type (e.g., a Debian package may be shipped by \texttt{debian} or its \texttt{ubuntu} derivative). Additionally, key-value pairs called \emph{qualifiers} can provide further context about a package, such as architecture, platform, or build metadata. Similar to a standard URL, qualifiers are appended to the basic \texttt{pURL} after a `\texttt{?}' character, with each key-value pair separated by an ampersand (`\texttt{\&}'). Lastly, a \texttt{subpath} may be included after the version and qualifiers, for instance, to indicate that the artifact only uses a function or a subpackage%
.

\subsubsection{System Package Data Exchange (SPDX)}

The System Package Data Exchange~\cite{SPDX:ONLINE} is an open standard for SBOM representation, initially introduced by the Linux Foundation. It comprises three main entities (see Figure~\ref{fig:spdx}). First, \texttt{packages} represent the installed software with their name, version, licenses, and other identifiers (such as the pURL). Then, \texttt{files} indicate the source or binary files that were found as part of the SCA process. Finally, \texttt{relationships} links files and packages together. Each entity is associated with a unique identifier.

\begin{figure}[tbp!]
\begin{lstlisting}[style=spdx]
"spdxVersion": "SPDX-2.3",
"dataLicense": "CC0-1.0",
"name": "alpine",
"creationInfo": {
 "licenseListVersion": "3.25",
 "creators": [
  "Organization: Anchore, Inc",
  "Tool: syft-1.12.2"
 ],
 "created": "2025-06-12T12:33:05Z"
 "packages": [{
   "(*@\color{black}name@*)": "binutils",
   "SPDXID": "SPDXRef-Package",
   "versionInfo": "2.40.2",
    "externalRefs": [{
     "referenceCategory": "PACKAGE-MANAGER",
     "referenceType": "purl",
     "referenceLocator": "pkg:deb/debian/binutils"
    }],
    ...
   }],
 "files": [{
  "fileName": "/bin/readelf",
  "SPDXID": "SPDXRef-File",
  ...
 }],
 "relationships": [{
  "spdxElementId": "SPDXRef-Package",
  "relationshipType": "CONTAINS",
  "relatedSpdxElement": "SPDXRef-File"
 }]
}
\end{lstlisting}%
    \caption{Minimal example of a SPDX file.}\label{fig:spdx}
\end{figure}

\subsection{Vulnerability Management}

Vulnerability management is the process of identifying, evaluating, and mitigating security vulnerabilities in software systems. In the context of SCA, it involves continuously monitoring third-party components for known vulnerabilities and updating them accordingly. Package managers have dedicated security teams responsible for monitoring and addressing Common Vulnerabilities and Exposures (CVEs). These CVEs may originate from external sources such as the MITRE corporation. %
Security trackers are systems that collect, analyze, and publish information about known vulnerabilities affecting packages in supported releases~\cite{Miranda:2021:TNSM}. Different Linux distributions have their own security trackers that offer an Application Programming Interface (API) for users to run queries against their vulnerability database~\cite{Lin:2023:ESE}.

\subsubsection{Mapping SBOMs to vulnerabilities}

SBOMs can be used to find vulnerabilities affecting the packages they describe. The process of mapping vulnerabilities is substantially different from creating the SBOM, however, many SCA tools~\cite{trivy,docker_scout,snyk,AWS:Inspector:web,AzureSBOM} support carrying out the two tasks at the same time. The mapping between an SBOM package identifier to CVEs is generally achieved through vendor-specific advisories or databases such as the NVD.
Usually, the tools that map packages to CVEs do not directly leverage online databases, but rather employ local copies and create custom views for faster lookup.

%% file: methodology.tex
\section{Overview}
\label{sec:dataset}

This section first discusses the motivation and goals of our work. It then introduces the methodology used for the evaluation, including the considered datasets and tools.

\paragraph{Motivation and Goals}

SBOMs have been promoted as a key mechanism to enhance software supply chain transparency and improve vulnerability management. Unfortunately, SBOM generation is often inconsistent across tools and ecosystems in practice~\cite{ODonoghue:2024:SCORED,PLATE:2023:ONLINE}. These discrepancies enable the risk of alleged compliance, where producers generate SBOMs that technically conform to standards but are ambiguous and not interoperable. This is especially the case for SBOMs generated for container images~\cite{Kawaguchi24}.

This study has three key objectives. First, we aim at examining the reasons and consequences of alleged compliance in such SBOMs. Second, we systematically target discrepancies in package reporting and vulnerability detection across different tools. Finally, we set out to provide actionable guidelines for SBOM generation to promote interoperability and reliability. We further support the latter with the release of a software that translates SBOMs between the different ``dialects'' employed by individual tools.

Achieving such objectives entails addressing several challenges. Discrepancies in reported packages and vulnerabilities are relatively straightforward to find, however, identifying their root causes is not at all trivial. To address this, we carry out a detailed analysis of both package identifiers and security trackers. In fact, OS packages require %
many details to be uniquely identified %
which are hard to correctly and consistently handle, unlike artifacts for programming languages. %

\begin{table}[tbp!]
    \def\arraystretch{1.125}
    \centering\footnotesize\sffamily
    \codefontsize
    \begin{tabular}{lrrr}
        \toprule
         & \multicolumn{2}{c}{\textbf{Packages}} \\
        \textbf{Dataset} & {Total}  & {Distinct (Avg)} & \textbf{Vulnerabilities} \\
        \cmidrule(lr){1-1} \cmidrule(lr){2-3} \cmidrule(lr){4-4}
        {All Debian}       & 63,461   & --         & 1,425     \\
        {All Alpine}       & 5,450    & --         & 113       \\
        {Top 20 Debian}    & 3,651    & 747  (192) & 261       \\
        {Top 20 Alpine}    & 741     & 442  (7)    & 78        \\
        \bottomrule
    \end{tabular}
    \caption{Considered datasets and their characteristics.}
    \label{tab:datasets}
\end{table}

\begin{table*}[tp!]
    \def\arraystretch{1.125}
    \centering\footnotesize\sffamily
    \codefontsize
    \begin{tabular}{lrlcccc}
        \toprule
        \textbf{Tool} & \textbf{Version} &\textbf{Vendor} & \textbf{Free}\,/\,\textbf{OSS} & \textbf{SBOM Output} & \scalebox{.95}[1.0]{\textbf{Vulnerability scan}} & \textbf{SBOM Input}\\
        \cmidrule(lr){1-1} \cmidrule(lr){2-3} \cmidrule(lr){4-7}
        Amazon Inspector~\cite{AWS:Inspector:web} & -- & Amazon           & \no & \yes & \yes & \partially \\ 
        Syft~\cite{syft}  & 0.77.0 & Anchore                                    & \yes & \yes & \no & \na \\
        Grype~\cite{grype} & 0.77.0 & Anchore                                   & \yes & \na & \yes & \yes \\
        Docker Scout~\cite{docker_scout} & 1.11.0 & Docker                      & \partially & \yes & \yes & \yes \\
        Artifact Analysis~\cite{GCloud} & -- & Google                     & \no & \yes & \yes & \no \\ 
        \texttt{sbom-tool}~\cite{SBOMTOOL:MICROSOFT:ONLINE} & 2.2.9 & Microsoft & \yes & \yes & \no & \na \\
        Defender for Cloud~\cite{AzureSBOM} & -- & Microsoft              & \no & \no & \yes & \no \\ 
        Trivy~\cite{trivy} & 0.50.2 & Trivy                                     & \yes & \yes & \yes & \yes \\
        \bottomrule
    \end{tabular}
    \caption{Software tools and cloud services considered in the evaluation. %
    The symbols %
    indicate whether the corresponding feature is \yes{} fully, \partially{} partially, \no{} not available, or \na{} not applicable (e.g., for individual tools that only target either SBOM generation or CVE scanning). Partial support means that the tool is free, but it is not available as open source software or supports formats other than SPDX for the SBOM output.} 
    \label{tab:tools}
\end{table*}

\paragraph{Datasets}

We employ different classes of datasets to evaluate the characteristics of different tools. The first class covers all the packages available in a distribution for a comprehensive evaluation. For this purpose, we considered the Debian \texttt{bookworm} release (as of May 16, 2025) and all the packages in Alpine \texttt{main} repository for its version \texttt{3.19}. %
Specifically, we created a container image with package manager state files that included all the packages in the respective distributions. %
The second class is represented by the most widely used images in Docker Hub.
For comparison purposes, we considered only the top 20 images that have both Debian and Alpine as base images and selected the last version deployed in 2024.

We leverage the content of the package manager state files (\texttt{dpkg/status} and \texttt{apk/db/installed}) to define our ground truth for the package information. Specifically, we utilized the CVE data from the Debian Security Tracker~\cite{DebianSecurityTracker} (commit \texttt{dbc2a94}) and Alpine Security Tracker~\cite{AlpineSecurityTracker} (website as of August 15, 2025). %
We followed the guidelines from the individual security trackers to define the ground truth for the CVEs. We employed a custom version of the SQL query used to display vulnerable packages~\cite{DST:2025:SORTING} so as to obtain all CVEs that apply to a given package. We only focused on CVEs that have been assigned by December 31, 2024 to better focus on the detection mechanisms of each tool, irrespective of the update cycle of their CVE databases~\cite{Miranda:2021:TNSM}. Table~\ref{tab:datasets} provides detailed information about the number of packages and vulnerabilities for each considered dataset.

\paragraph{Considered Tools}

We take a black-box approach and consider diverse tools ranging from free and open source software to cloud services. Specifically, we selected the most widely used software-based solutions~\cite{RedHat:2024:web} (namely, Grype, Syft, Docker Scout, and Trivy) and products offered by major cloud providers (i.e., Google, Amazon Web Services, and Azure). Table~\ref{tab:tools} provides the details on the considered tools.

Note that some tools are very specific, in the sense that they either generate SBOMs (i.e., Syft) or carry out vulnerability scanning (i.e., Grype and Defender for Cloud). Most tools can do both but in some cases only accept the container image as input instead of the SBOM (i.e., Artifact Analysis and Defender for Cloud). Because of this, we use the vendor in Table~\ref{tab:tools} to identify a set of separate tools jointly utilized or an individual solution that performs end-to-end analysis (i.e., both SBOM generation and vulnerability scanning) for convenience in the rest of the paper.

\paragraph{Evaluation}

Our evaluation broadly targets the ``usability'' of SBOMs within SCA, with particular emphasis on interoperability. In fact, SCA relies on SBOMs for vulnerability analysis; the latter could be done with a tool that is different from what has generated the SBOM in the first place. This is especially the case for organizations obtaining SBOMs from closed-source components from third parties. The practical issues caused by SBOM discrepancies include failures in the mapping between the packages identified by a given tool and the CVEs recognized by another one, leading to several cases of false negatives that ultimately make an artifact more vulnerable.

Our evaluation consists of two separate studies. The first one (in Section~\ref{sec:compliance}) targets \emph{SBOM compliance}, namely, how the SBOM files created by the considered tools adhere to the pURL and SPDX specifications. 
The evaluation especially focuses on understanding their issues and root causes.
The second part (in Section~\ref{sec:accuracy}) focuses on the \emph{detection accuracy} of the considered tools. Specifically, we considered the accuracy in detecting both packages and vulnerabilities. In this case, we report the true (false) positive\,/\,negatives in addition to qualitative results.

%% file: compliance.tex
\section{SBOM Compliance} %
\label{sec:compliance}

This section explores the issue of SBOM compatibility. It begins by considering pURLs as the basis for accurate package identification. Then it delves into the peculiar use of the SPDX format by (some of) the considered tools.

\subsection{Package URLs}
\label{sec:purl}

The pURL standard aims to be the universal identifier of an SBOM package -- indeed, all the tools we considered rely only on it. %
For this reason, we focused on whether the pURLs in SBOMs generated by the different tools are actually interchangeable.

\begin{figure}[tbp!]
    \centering
    \hspace{-7.25pt}%
    \subcaptionbox{\label{fig:jaccard-purl:debian:all}}{%
        \centering%
        \heatmap{%
            Amazon,Anchore,Docker,Gcloud,Microsoft,Trivy%
        }{%
            Amazon,Anchore,Docker,Gcloud,Microsoft,Trivy%
        }{%
            {1.00,0.00,0.00,0.00,0.00,0.00},%
            {0.00,1.00,0.00,0.16,0.00,0.00},%
            {0.00,0.00,1.00,0.00,0.00,0.00},%
            {0.00,0.16,0.00,1.00,0.00,0.00},%
            {0.00,0.00,0.00,0.00,1.00,0.00},%
            {0.00,0.00,0.00,0.00,0.00,1.00},%
        }%
    }%
    \subcaptionbox{\label{fig:jaccard-purl:debian:top20}}{%
        \centering%
        \heatmap{%
            Amazon,Anchore,Docker,Gcloud,Microsoft,Trivy%
        }{%
            Amazon,Anchore,Docker,Gcloud,Microsoft,Trivy%
        }{%
            {1.00,0.00,0.00,0.00,0.00,0.00},%
            {0.00,1.00,0.00,0.08,0.00,0.00},%
            {0.00,0.00,1.00,0.00,0.00,0.00},%
            {0.00,0.08,0.00,1.00,0.00,0.00},%
            {0.00,0.00,0.00,0.00,1.00,0.00},%
            {0.00,0.00,0.00,0.00,0.00,1.00},%
        }%
    }%
    \caption{Jaccard index for the pURLs in the Debian \subref{fig:jaccard-purl:debian:all} All and \subref{fig:jaccard-purl:debian:top20} Top 20 datasets.}\label{fig:jaccard-purls}
\end{figure}

First, we compared the similarity of these pURLs for the Debian datasets by using the Jaccard index~\cite{Niwattanakul:2013:IMECS} (between 0 and 1, the higher the more similar). Figure~\ref{fig:jaccard-purls} clearly shows that the pURLs generated by the considered tools are substantially different from each other, with only a minor overlap occurring between Anchore and Google. We found similar results in the Alpine datasets, so we do not report them for conciseness.

Then, we delved into such differences by analyzing the discrepancies between the pURLs generated by the considered tools and the specification. Our analysis revealed several issues, summarized in Table~\ref{fig:tab:purl-python3-magic} and discussed next.

\begin{table}[tbp!]
    \newcommand{\high}[1]{\textcolor{aaltoRed}{#1}}
    \newcommand{\medium}[1]{\textcolor{aaltoOrange}{#1}}
    \newcommand{\low}[1]{\textcolor{aaltoPurple}{#1}}
    \def\arraystretch{1.125}
    \setlength{\tabcolsep}{4.25pt}
    \centering\footnotesize\sffamily
    \codefontsize
    \begin{tabular}{p{1.625cm}p{6.65cm}}
        \toprule
        \textbf{Tool}      &   \textbf{pURL}   \\
        \cmidrule(lr){1-1} \cmidrule(lr){2-2}
        Amazon & \texttt{\low{pkg:dpkg}/\low{python3-magics++}@\low{1.5.8-1}?arch=\medium{AMD64}\& \high{epoch=1}\&\high{upstream=python3-magics++-1.5.8-1.src.dpkg}}
        \\
        Anchore & \texttt{pkg:deb/debian/python3-magics\%2B\%2B@2:1.5.8-1? arch=amd64\&\high{upstream=magics-python}\&distro=\medium{debian-12}} \\
        Google & \texttt{pkg:deb/debian/python3-magics\%2B\%2B@\low{2\%3A1.5.8-1} ?arch=amd64\&distro=\medium{debian-12} \&\high{upstream=magics-python}} \\
        Microsoft & \texttt{pkg:deb/debian/\low{python3-magics++}@2:1.5.8-1} \\
        Scout & \texttt{pkg:deb/debian/python3-magics\%2B\%2B@2:1.5.8-1? \high{os\_version=12}\&\high{os\_name=debian}\&\high{os\_distro=bookworm}} 
        \\
        Trivy & \texttt{pkg:deb/debian/python3-magics\%2B\%2B@\low{1.5.8-1}? arch=amd64\&distro=\medium{debian-12.11}\&\high{epoch=2}}
        \\
        \cmidrule(lr){1-1} \cmidrule(lr){2-2}
        Reference~\cite{PURL:2025:TYPES} & \texttt{pkg:deb/debian/python3-magics\%2B\%2B@2:1.5.8-1? arch=amd64\&distro=bookworm}

        \\
        \bottomrule
    \end{tabular}
    \caption{pURLs for the \texttt{python3-magics++} package from SBOMs generated by different tools and their issues with respect to the specification. Text in \medium{orange} means that an optional qualifier is correct but the value is incorrect, \high{red} indicates that an optional qualifier is incorrect, \low{purple} highlights mandatory pURL parameters that are not compliant. The last row shows a pURL that complies with the specifications as a reference.}
    \label{fig:tab:purl-python3-magic}
\end{table}

\subsubsection{Invalid Format}
\label{issue:purl:invalid}

The first and most important outcome of our analysis is that \emph{no tool fully respects the pURL format}. In fact, we found issues affecting both optional qualifiers and mandatory parameters. Amazon, for instance, uses a non-standard \texttt{dpkg} type that is not recognized by any other tool. Another widespread problem is the use of the \texttt{epoch} in the package version. The specification~\cite{purl:spec} clearly states that the epoch qualifier does not apply to Debian. Nevertheless, Amazon and Trivy all use the epoch in the package version. Optional qualifiers also have well-defined keys and value formats. For instance, valid qualifiers for Debian are \texttt{distro} and \texttt{arch}, their values are lowercase, and distro should represent the codename of the operating system. Scout uses non-standard qualifiers, Trivy uses a different distro format, and Amazon uses uppercase arch values. Finally, string encoding is also a cause of mismatch. Amazon and Microsoft do not encode package names as the other tools correctly do~\cite{PURL:2025:TYPES}.

\subsubsection{Incomplete Data}
\label{issue:purl:incomplete}

The main problem pURL tries to solve is providing a unique string that contains all the necessary information to identify an artifact~\cite{ecma:purl}. Unfortunately, we found that some pURLs lack critical information and do not allow for complete identification of a package. An example of this is the Microsoft pURL, which does not include information about the architecture or distribution of the package. For instance, the kernel vulnerability \texttt{CVE-2016-2143} only applies to the \texttt{s390} architecture. Another example is \texttt{CVE-2023-29383}, which targets the \texttt{shadow} package in Debian. The Debian \texttt{bullseye} release is vulnerable, while \texttt{bookworm} and \texttt{trixie} are not.

\subsubsection{Incorrect Information}
\label{issue:purl:incorrect}

We found that Amazon pURLs often exhibit incorrect upstream information. The upstream is used to track the originating package, and it reflects the source package in Debian. Amazon always uses the package name and version as upstream, which is incorrect. For instance, the following pURL:
\begin{center}
    \observationfontsize
    \texttt{pkg:dpkg/libelf1@0.188-2.1?arch=AMD64\&epoch=0\\\&upstream=libelf1-0.188-2.1.src.dpkg}
\end{center}
lists \texttt{libelf1-0.188-2.1} as the upstream, while the correct value is instead \texttt{elfutils} (as reported in the package manager status file). The upstream field is heavily used to search for vulnerabilities and these inaccuracies can cause false positives\,/\,negatives. %
Moreover, Amazon pURLs often report incorrect version information. %

\begin{observation}
    Tools do not employ the same pURLs and none of them respects the standard. Some tools generate incomplete or incorrect pURLs.
\end{observation}

\begin{figure*}[tbp!]
    \centering
    \subcaptionbox{\label{fig:jaccard-ids:debian:all}}{%
        \centering%
        \heatmap{%
            Amazon,Anchore,Docker,Gcloud,Microsoft,Trivy%
        }{%
            Amazon,Anchore,Docker,Gcloud,Microsoft,Trivy%
        }{%
            {1.00,0.54,0.68,0.54,0.54,0.54},%
            {0.54,1.00,0.82,1.00,1.00,1.00},%
            {0.68,0.82,1.00,0.82,0.82,0.82},%
            {0.54,1.00,0.82,1.00,1.00,1.00},%
            {0.54,1.00,0.82,1.00,1.00,1.00},%
            {0.54,1.00,0.82,1.00,1.00,1.00},%
        }%
    }%
    \subcaptionbox{\label{fig:jaccard-ids:debian:top20}}{%
        \centering%
        \heatmap{%
            Amazon,Anchore,Docker,Gcloud,Microsoft,Trivy%
        }{%
            Amazon,Anchore,Docker,Gcloud,Microsoft,Trivy%
        }{%
            {1.00,0.61,0.81,0.61,0.61,0.61},%
            {0.61,1.00,0.76,1.00,1.00,1.00},%
            {0.81,0.76,1.00,0.76,0.76,0.76},%
            {0.61,1.00,0.76,1.00,1.00,1.00},%
            {0.61,1.00,0.76,1.00,1.00,1.00},%
            {0.61,1.00,0.76,1.00,1.00,1.00},%
        }%
    }%
    \subcaptionbox{\label{fig:jaccard-ids:alpine:all}}{%
        \centering%
        \heatmap{%
            Amazon,Anchore,Docker,Gcloud,Microsoft,Trivy%
        }{%
            Amazon,Anchore,Docker,Gcloud,Microsoft,Trivy%
        }{%
            {1.00,1.00,1.00,1.00,1.00,1.00},%
            {1.00,1.00,1.00,1.00,1.00,1.00},%
            {1.00,1.00,1.00,1.00,1.00,1.00},%
            {1.00,1.00,1.00,1.00,1.00,1.00},%
            {1.00,1.00,1.00,1.00,1.00,1.00},%
            {1.00,1.00,1.00,1.00,1.00,1.00},%
        }%
    }%
    \subcaptionbox{\label{fig:jaccard-ids:alpine:top20}}{%
        \centering%

        \heatmap{%
            Amazon,Anchore,Docker,Gcloud,Microsoft,Trivy%
        }{%
            Amazon,Anchore,Docker,Gcloud,Microsoft,Trivy%
        }{%
            {1.00,0.81,0.95,0.81,0.77,0.81},%
            {0.81,1.00,0.85,1.00,0.95,1.00},%
            {0.95,0.85,1.00,0.85,0.81,0.85},%
            {0.81,1.00,0.85,1.00,0.95,1.00},%
            {0.77,0.95,0.81,0.95,1.00,0.95},%
            {0.81,1.00,0.85,1.00,0.95,1.00},%
        }
    }%
    \caption{Jaccard index for the package identifiers in the: Debian \subref{fig:jaccard-ids:debian:all} All and \subref{fig:jaccard-ids:debian:top20} Top 20 datasets; Alpine \subref{fig:jaccard-ids:alpine:all} All and \subref{fig:jaccard-ids:alpine:top20} Top 20 datasets.}\label{fig:jaccard-ids}
\end{figure*}

\subsection{SPDX Format}
\label{sec:format}

The SPDX \emph{package} entity contains a pURL in addition to other data about the package. In principle, none of such other data in the package should be used to search for CVEs, especially if the fields are optional. However, we found that it is not the case.

 \begin{table}[tbp!]
    \def\arraystretch{1.125}
    \setlength{\tabcolsep}{4.25pt}
    \centering\footnotesize\sffamily
    \codefontsize
    \begin{tabular}{lll}
        \toprule
        \textbf{Tool} & \textbf{Qualifiers} & \textbf{Unused}   \\
        \cmidrule(lr){1-1} \cmidrule(lr){2-3}
        Trivy   & \texttt{arch}, \texttt{distro}, \texttt{epoch} & \texttt{arch}, \texttt{distro}, \texttt{epoch} \\
        Anchore & \texttt{arch}, \texttt{distro}, \texttt{upstream} & \texttt{arch}\\
        Docker  & \texttt{os\_distro}, \texttt{os\_version}, \texttt{os\_name} & \texttt{os\_distro}\\
        \bottomrule
        \end{tabular}
        \caption{Table of optional pURL qualifiers by tool.}
        \label{tab:purl_qualifiers}
\end{table}

\subsubsection{pURL Not Used As Package Identifier}
\label{issue:format:nopurl}

We first checked if pURLs are actually used to index packages. For this purpose, we took the SBOM of a Debian container and selected a target package to test. For that package, we edited the qualifier in the pURL and investigated which changes (for that qualifier) affected the number of discovered vulnerabilities. Table~\ref{tab:purl_qualifiers} reports these qualifiers, as well as those that are not used in practice. We noticed that no change in the qualifier affected Trivy's findings. This actually occurs as Trivy ignores all of them, since it relies on storing package information in the optional \texttt{sourceInfo} field with a specific%
\begin{center}
    \observationfontsize
    \texttt{built package from: <package-name> <package-version>}
\end{center}
The package version has a different format than the version in the pURL (see Figure~\ref{fig:trivy} in appendix for an example). In fact, the package version of the \texttt{sourceInfo} optional fields has to be in the form: \texttt{epoch:version}. The package name is instead set to the Debian upstream name that corresponds to the Debian source package.
We repeated the same experiment, but using only the package name and version. We noticed that Grype employs version information from the SBOM package instead of that inside the pURL to index CVEs. We obtained the same results with an Alpine container, thereby supporting our findings.

\subsubsection{Reliance on Optional Fields and Custom Entries}
\label{issue:format:fields}

Another issue we discovered relates to the usage of optional fields. In particular, we noticed that Docker Scout requires the optional field \texttt{primaryPackagePurpose} to be set in each package, otherwise it fails to parse the SBOM. Similarly, we noticed that Trivy requires the SBOM to include a custom package entry with the following settings:
\begin{center}
    \observationfontsize
    \texttt{"attributionTexts": ["Class: os-pkgs", 
    "Type: <os-name>"],}\\
    \texttt{primaryPackagePurpose}=\texttt{OPERATING-SYSTEM}
\end{center}
Without that package, Trivy is unable to discover any vulnerability because it cannot determine which OS the package belongs to. 

\begin{observation}
    Some tools do not use pURLs to uniquely identify packages and rely on other SPDX parameters or optional fields.
\end{observation}

%% file: accuracy.tex
\section{Detection Accuracy} %
\label{sec:accuracy}

This section characterizes the accuracy of the different tools in detecting vulnerable packages and the source CVEs.

\subsection{Vulnerable Packages}
\label{sec:accuracy:packages}

\begin{table*}[tb!]
    \centering
    \subcaptionbox{\label{tab:pkgs}}{%
        \def\arraystretch{1.125}
        \setlength{\tabcolsep}{4.25pt}
        \centering\footnotesize\sffamily
        \codefontsize
        \begin{tabular}{clrrrrrrrr}
            \toprule
             & & \multicolumn{4}{c}{\textbf{All packages}} & \multicolumn{4}{c}{\textbf{Excluding kernel packages}} \\
             \textbf{Dataset} & \textbf{Tool} & {\# Pkgs} & {Vuln.} & {CVEs} & {Dups}
                                                 & {\# Pkgs} & {Vuln.} & {CVEs} & {Dups} \\
            \cmidrule(lr){1-1} \cmidrule(lr){2-2} \cmidrule(lr){3-6} \cmidrule(lr){7-10} 
            \multirow{6}{1cm}{\raggedleft\textbf{Debian All}}
            \def\arraystretch{1}
            & Amazon      &          80608 &                        49 &        576 &                  574 &          80527 &                        47 &        194 &                  193 \\
            & Anchore     &          63461 &                      3998 &      30577 &                 2725 &          63388 &                      3966 &      16433 &                 2258 \\
            & Docker      &          82619 &                       507 &       1370 &                 1292 &          82538 &                       507 &       1370 &                 1292 \\
            & Gcloud      &          63461 &                       853 &       2675 &                 2468 &          63388 &                       853 &       2675 &                 2468 \\
            & Microsoft   &          69006 &                         9 &        203 &                  203 &          68933 &                         9 &        203 &                  203 \\
            & Trivy       &          63461 &                      3975 &      35650 &                 2922 &          63388 &                      3933 &      18016 &                 2473 \\
            \cmidrule(lr){1-1} \cmidrule(lr){2-2} \cmidrule(lr){3-6} \cmidrule(lr){7-10} 
            \multirow{6}{1cm}{\raggedleft\textbf{Alpine All}} 
            & Amazon      &           5561 &                        42 &        100 &                  100 &           5437 &                        40 &         98 &                   98 \\
            & Anchore     &           5450 &                       553 &       1232 &                  157 &           5330 &                       551 &       1229 &                  157 \\
            & Docker      &           5561 &                        43 &        105 &                  105 &           5437 &                        41 &        103 &                  103 \\
            & Gcloud      &           5450 &                        40 &         99 &                   99 &           5330 &                        39 &         98 &                   98 \\
            & Microsoft   &           5518 &                        41 &         99 &                   99 &           5398 &                        41 &         98 &                   98 \\
            & Trivy       &           5450 &                       357 &        936 &                   99 &           5330 &                       355 &        934 &                   99 \\
            \cmidrule(lr){1-1} \cmidrule(lr){2-2} \cmidrule(lr){3-6} \cmidrule(lr){7-10} 
            \multirow{6}{1cm}{\raggedleft\textbf{Debian Top 20}}
            & Amazon      &           5063 &                        31 &       2692 &                 1477 &           5047 &                        27 &         75 &                   64 \\
            & Anchore     &           3651 &                       331 &       1276 &                  263 &           3643 &                       331 &       1276 &                  263 \\
            & Docker      &           5071 &                        81 &        221 &                  207 &           5055 &                        81 &        221 &                  207 \\
            & Gcloud      &           3651 &                        96 &        347 &                  263 &           3643 &                        96 &        347 &                  263 \\
            & Microsoft   &           3652 &                         7 &         11 &                   11 &           3644 &                         7 &         11 &                   11 \\
            & Trivy       &           3651 &                       333 &       4547 &                 1747 &           3643 &                       328 &       1243 &                  263 \\
            \cmidrule(lr){1-1} \cmidrule(lr){2-2} \cmidrule(lr){3-6} \cmidrule(lr){7-10} 
            \multirow{6}{1cm}{\raggedleft\textbf{Alpine Top 20}}
            & Amazon      &            897 &                        12 &         15 &                   11 &            890 &                        12 &         15 &                   11 \\
            & Anchore     &            741 &                        66 &        113 &                   51 &            740 &                        66 &        113 &                   51 \\
            & Docker      &            897 &                        11 &         13 &                    9 &            890 &                        11 &         13 &                    9 \\
            & Gcloud      &            741 &                         8 &         10 &                    7 &            740 &                         8 &         10 &                    7 \\
            & Microsoft   &            785 &                         9 &         11 &                   11 &            784 &                         9 &         11 &                   11 \\
            & Trivy       &            741 &                        21 &         27 &                   11 &            740 &                        21 &         27 &                   11 \\
            \bottomrule
        \end{tabular}
    }%
    \hspace{2em}%
    \subcaptionbox{\label{tab:average_cves}}{%
        \def\arraystretch{1.125}
        \setlength{\tabcolsep}{4.25pt}
        \centering\footnotesize\sffamily
        \codefontsize
        \begin{tabular}{clr}
            \toprule
            \textbf{Dataset} & \textbf{Tool} & \textbf{Duplicate CVEs}\\
            \cmidrule(lr){1-1} \cmidrule(lr){2-2} \cmidrule(lr){3-3}
            \multirow{6}{1cm}{\raggedleft\textbf{Debian All}}
            & Amazon    & 1.00 (0.00) \\
            & Anchore   & 7.28 (0.31) \\
            & Microsoft & 1.00 (0.00) \\
            & Docker    & 1.06 (0.01) \\
            & Trivy     & 7.27 (0.27) \\
            & Gcloud    & 1.05 (0.01) \\
            \cmidrule(lr){1-1} \cmidrule(lr){2-2} \cmidrule(lr){3-3}
            \multirow{6}{1cm}{\raggedleft\textbf{Alpine All}} 
            & Amazon    & 1.00 (0.00) \\
            & Anchore   & 7.83 (0.61) \\
            & Microsoft & 1.00 (0.00) \\
            & Docker    & 1.00 (0.00) \\
            & Trivy     & 9.43 (0.72) \\
            & Gcloud    & 1.00 (0.00) \\
            \cmidrule(lr){1-1} \cmidrule(lr){2-2} \cmidrule(lr){3-3}
            \multirow{6}{1cm}{\raggedleft\textbf{Debian Top 20}}
            & Amazon    & 1.17 (0.05) \\
            & Anchore   & 4.85 (0.39) \\
            & Microsoft & 1.00 (0.00) \\
            & Docker    & 1.07 (0.02) \\
            & Trivy     & 4.73 (0.37) \\
            & Gcloud    & 1.36 (0.04) \\
            \cmidrule(lr){1-1} \cmidrule(lr){2-2} \cmidrule(lr){3-3}
            \multirow{6}{1cm}{\raggedleft\textbf{Alpine Top 20}}
            & Amazon    & 1.36 (0.15) \\
            & Anchore   & 2.22 (0.37) \\
            & Microsoft & 1.00 (0.00) \\
            & Docker    & 1.44 (0.18) \\
            & Trivy     & 2.45 (0.65) \\
            & Gcloud    & 1.43 (0.20)  \\
            \bottomrule
        \end{tabular}
    }%
    \caption{\subref{tab:pkgs} Comparison of vulnerability scanning tools with and without kernel packages. The subcolumns indicate Packages, Vulnerable Packages, total CVEs, Distinct number of CVEs. \subref{tab:average_cves} Average number of packages with the same set of CVEs in the different datasets.}\label{tab:vulnerabilities}

\end{table*}

Vulnerable packages are those affected by one or more security issues described by a CVE (Table~\ref{tab:vulnerabilities}).

\subsubsection{Total Packages Differ}

As a preliminary step, we looked into the total number of packages recognized by the different tools. The numbers are actually not the same as one would instead expect. In particular, Amazon and Docker report remarkably higher numbers than the rest of the tools. In particular, Anchore, Google, and Trivy detect the same amount of packages, matching those actually present in the datasets (see Table~\ref{tab:datasets}). Instead, Microsoft reports up to 10\% additional packages compared to the correct value. These issues are more noticeable in the \emph{Debian all} and \emph{Alpine all} datasets because there are no duplicate packages with different versions therein.

We then focused on these additional packages found by Amazon, Docker, and Microsoft. First, we computed the Jaccard index for the package identifiers in the different datasets, shown in Figure~\ref{fig:jaccard-ids}. Then, we compared the additional packages against those listed in the status
file of the package managers. Microsoft detects a relatively limited number of additional packages as it considers the information of every layer of the container, whereas the others analyze the outcome of applying all the layers. Amazon and Docker, instead, include the actual packages in their SPDX output but also generate an additional entry when the package name differs from the source package name. For instance, containers usually include the \texttt{login} binary (package), which is only one of those produced by the \texttt{shadow} source package (e.g., \texttt{uidmap}).
Such a behavior overstates the total number of packages and motivates a deeper analysis of the CVEs. %
Interestingly, only the \textit{Alpine all} dataset has the same package identifiers (although the total number of packages differs due to duplicate entries). That is because in Alpine, there is always a binary package with the same name as the source package.

\begin{observation}
    Tools roughly find the same packages but report them differently: some only indicate the package name, others duplicate entries when package and source names are not the same. 
\end{observation}

\begin{figure*}[tbp!]
    \centering
    \subcaptionbox{\label{fig:jaccard-vulnerabilities:debian:all}}{%
        \heatmap[-20]{%
            Amazon,Anchore,Docker,Gcloud,Microsoft,Trivy%
        }{%
            Amazon,Anchore,Docker,Gcloud,Microsoft,Trivy%
        }{%
            {1.00,0.01,0.05,0.00,0.00,0.01},%
            {0.01,1.00,0.06,0.54,0.00,0.86},%
            {0.05,0.06,1.00,0.05,0.00,0.07},%
            {0.00,0.54,0.05,1.00,0.00,0.52},%
            {0.00,0.00,0.00,0.00,1.00,0.00},%
            {0.01,0.86,0.07,0.52,0.00,1.00},%
        }%
    }%
    \hfill%
    \subcaptionbox{\label{fig:jaccard-vulnerabilities:alpine:all}}{%
        \heatmap[-20]{%
            Amazon,Anchore,Docker,Gcloud,Microsoft,Trivy%
        }{%
            Amazon,Anchore,Docker,Gcloud,Microsoft,Trivy%
        }{%
            {1.00,0.07,1.00,0.54,0.55,0.11},%
            {0.07,1.00,0.07,0.12,0.12,0.56},%
            {1.00,0.07,1.00,0.54,0.55,0.11},%
            {0.54,0.12,0.54,1.00,0.93,0.20},%
            {0.55,0.12,0.55,0.93,1.00,0.20},%
            {0.11,0.56,0.11,0.20,0.20,1.00},%
        }%
    }
    \hfill%
    \subcaptionbox{\label{fig:jaccard-vulnerabilities:debian:top20}}{%
        \heatmap[-20]{%
            Amazon,Anchore,Docker,Gcloud,Microsoft,Trivy%
        }{%
            Amazon,Anchore,Docker,Gcloud,Microsoft,Trivy%
        }{%
            {1.00,0.03,0.27,0.03,0.08,0.03},%
            {0.03,1.00,0.09,0.83,0.04,0.98},%
            {0.27,0.09,1.00,0.09,0.03,0.09},%
            {0.03,0.83,0.09,1.00,0.04,0.83},%
            {0.08,0.04,0.03,0.04,1.00,0.04},%
            {0.03,0.98,0.09,0.83,0.04,1.00},%
        }%
    }
    \hfill%
    \subcaptionbox{\label{fig:jaccard-vulnerabilities:alpine:top20}}{%
        \heatmap[-20]{%
            Amazon,Anchore,Docker,Gcloud,Microsoft,Trivy%
        }{%
            Amazon,Anchore,Docker,Gcloud,Microsoft,Trivy%
        }{%
            {1.00,0.11,0.92,0.22,0.44,0.32},%
            {0.11,1.00,0.10,0.15,0.21,0.21},%
            {0.92,0.10,1.00,0.24,0.39,0.28},%
            {0.22,0.15,0.24,1.00,0.71,0.48},%
            {0.44,0.21,0.39,0.71,1.00,0.67},%
            {0.32,0.21,0.28,0.48,0.67,1.00},%
        }%
    }
    \caption{Jaccard index for the vulnerable packages in the: Debian \subref{fig:jaccard-vulnerabilities:debian:all} All and \subref{fig:jaccard-vulnerabilities:debian:top20} Top 20 datasets; Alpine \subref{fig:jaccard-vulnerabilities:alpine:all} All and \subref{fig:jaccard-vulnerabilities:alpine:top20} Top 20 datasets.}\label{fig:jaccard-vulnerabilities}
\end{figure*}

\subsubsection{Source-Binary Mismatch}
\label{sec:source-binary-mismatch}

We have just seen that most tools essentially recognize the same packages; however, those they report as vulnerable often significantly differ between each other even in such a case (see Table~\ref{tab:pkgs}). Also, here we consider the package identifier of the vulnerable packages, illustrated in Figure~\ref{fig:jaccard-vulnerabilities}, to better understand the reason behind these issues. %
The results show several interesting patterns. First, Trivy and Anchore report a similar number of vulnerable packages in Debian but not in Alpine; Amazon shows some overlap with Docker in Alpine, while Google, Anchore, and Trivy have similar vulnerable packages in Debian Top 20.

Following our earlier analysis, the reason is how tools handle vulnerabilities with respect to binary and source packages. %
Trivy in particular does not rely on the pURL of each package, but instead leverages custom fields derived from the metadata of the upstream (as explained in Section~\ref{sec:compliance}). We now examine the number of packages with the same CVEs in Table~\ref{tab:average_cves} to better characterize multiple occurrences of vulnerable packages. The results highlight that four tools do not have duplicate source packages with the same CVEs. %

\begin{observation}
Amazon, Docker, Google, and Microsoft only use source package information to search for CVEs. 
\end{observation}

\begin{figure*}[tbp!]
    \centering
    \includegraphics[scale=1.0]{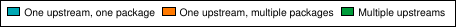}\\
    \subcaptionbox{\label{fig:package-type-vulnerabilities:debian:top20}}{%
        \plotfont
        \begin{tikzpicture}
            \begin{axis}[
                xscale=.6,yscale=.3,
                width=7cm,
                height=8cm,
                ybar stacked,
                ymin=0, ymax=100,
                bar width=3em,
                ylabel={CVEs per package (\%)},
                ylabel style = {yshift=-2.5em,xshift=3.125em},
                symbolic x coords={Amazon, Anchore, Docker, Gcloud, Microsoft,Trivy},
                xtick=data,
                xticklabel style={rotate=30, anchor=east, xshift=-.05em, yshift=-1em},
                yticklabel={\pgfmathprintnumber[assume math mode=true]{\tick}},
                enlarge x limits={abs=10pt},
                tick label style = {font=\sffamily},
                every axis label = {font=\sffamily},
                legend style = {font=\sffamily},
                label style = {font=\sffamily},
            ]
            \addplot [fill=aaltoTurquoise] coordinates {(Amazon,82.81) (Anchore,21.67) (Microsoft,100.00) (Docker,93.72) (Trivy,21.67) (Gcloud,75.39)};  %
            \addplot [fill=aaltoOrange] coordinates {(Amazon,17.19) (Anchore,78.33) (Microsoft,0.00) (Docker,6.28) (Trivy,78.33) (Gcloud,24.61)};%
            \addplot [fill=aaltoGreen] coordinates {(Amazon, 0) (Anchore, 0) (Docker, 0) (Gcloud, 0) (Microsoft, 0) (Trivy, 0)};%
            \end{axis}
        \end{tikzpicture}
    }%
    \hfill
    \subcaptionbox{\label{fig:package-type-vulnerabilities:debian:all}}{%
        \plotfont
        \begin{tikzpicture}
            \begin{axis}[
                xscale=.6,yscale=.3,
                width=7cm,
                height=8cm,
                ybar stacked,
                ymin=0, ymax=100,
                bar width=3em,
                ylabel={CVEs per package (\%)},
                ylabel style = {yshift=-2.5em,xshift=3.125em},
                symbolic x coords={Amazon, Anchore, Microsoft, Docker, Trivy, Gcloud},
                xtick=data,
                xticklabel style={rotate=30, anchor=east, xshift=-.05em, yshift=-1em},
                yticklabel={\pgfmathprintnumber[assume math mode=true]{\tick}},
                enlarge x limits={abs=10pt},
                tick label style = {font=\sffamily},
                every axis label = {font=\sffamily},
                legend style = {font=\sffamily},
                label style = {font=\sffamily},
            ]
            \addplot [fill=aaltoTurquoise] coordinates {(Amazon,99.48) (Anchore,25.07) (Microsoft,100.00) (Docker,94.27) (Trivy,27.21) (Gcloud,96.1)}; %
            \addplot [fill=aaltoOrange] coordinates {(Amazon,0.00) (Anchore,70.59) (Microsoft,0.00) (Docker,2.48) (Trivy,69.51) (Gcloud,0.00) }; %
            \addplot [fill=aaltoGreen] coordinates {(Amazon,0.52) (Anchore,4.34) (Microsoft,0.00) (Docker,3.25) (Trivy,3.28) (Gcloud,3.9)}; %

            \end{axis}
        \end{tikzpicture}
    }%
    \hfill
    \subcaptionbox{\label{fig:package-type-vulnerabilities:alpine:top20}}{%
        \plotfont
        \begin{tikzpicture}
            \begin{axis}[
                xscale=.6,yscale=.3,
                width=7cm,
                height=8cm,
                ybar stacked,
                ymin=0, ymax=100,
                bar width=3em,
                ylabel={CVEs per package (\%)},
                ylabel style = {yshift=-2.5em,xshift=3.125em},
                symbolic x coords={Amazon, Anchore, Microsoft, Docker, Trivy, Gcloud},
                xtick=data,
                xticklabel style={rotate=30, anchor=east, xshift=-.05em, yshift=-1em},
                yticklabel={\pgfmathprintnumber[assume math mode=true]{\tick}},
                enlarge x limits={abs=10pt},
                tick label style = {font=\sffamily},
                every axis label = {font=\sffamily},
                legend style = {font=\sffamily},
                label style = {font=\sffamily},
            ]
                \addplot [fill=aaltoTurquoise] coordinates {(Amazon,63.64) (Anchore,62.75) (Microsoft,100.00) (Docker,55.56) (Trivy,45.45) (Gcloud,57.14)}; %
                \addplot [fill=aaltoOrange] coordinates {(Amazon,36.36) (Anchore,35.29) (Microsoft,0.00) (Docker,44.44) (Trivy,54.55) (Gcloud,42.86)}; %
                \addplot [fill=aaltoGreen] coordinates {(Amazon,0.00) (Anchore,1.96) (Microsoft,0.00) (Docker,0.00) (Trivy,0.00) (Gcloud,0.00)}; %

            \end{axis}
        \end{tikzpicture}
    }%
    \hfill
    \subcaptionbox{\label{fig:package-type-vulnerabilities:alpine:all}}{%
        \plotfont
        \begin{tikzpicture}
            \begin{axis}[
                xscale=.6,yscale=.3,
                width=7cm,
                height=8cm,
                ybar stacked,
                ymin=0, ymax=100,
                bar width=3em,
                ylabel={CVEs per package (\%)},
                ylabel style = {yshift=-2.5em,xshift=3.125em},
                symbolic x coords={Amazon, Anchore, Microsoft, Docker, Trivy, Gcloud},
                xtick=data,
                xticklabel style={rotate=30, anchor=east, xshift=-.05em, yshift=-1em},
                yticklabel={\pgfmathprintnumber[assume math mode=true]{\tick}},
                enlarge x limits={abs=10pt},
                tick label style = {font=\sffamily},
                every axis label = {font=\sffamily},
                legend style = {font=\sffamily},
                label style = {font=\sffamily},
            ]
                \addplot [fill=aaltoTurquoise] coordinates {(Amazon,100.00) (Anchore,17.2) (Microsoft,100.00) (Docker,100.00) (Trivy,1.01) (Gcloud,100.00)}; %
                \addplot [fill=aaltoOrange] coordinates {(Amazon,0.00) (Anchore,82.17) (Microsoft,0.00) (Docker,0.00) (Trivy,98.99) (Gcloud,0.00)}; %
                \addplot [fill=aaltoGreen] coordinates {(Amazon,0.00) (Anchore,0.64) (Microsoft,0.00) (Docker,0.00) (Trivy,0.00) (Gcloud,0.00)}; %
            \end{axis}
        \end{tikzpicture}
    }%
    \caption{Types of CVEs per package in: Debian \subref{fig:package-type-vulnerabilities:debian:all} All and \subref{fig:package-type-vulnerabilities:debian:top20} Top 20 datasets; Alpine \subref{fig:package-type-vulnerabilities:alpine:all} All and \subref{fig:package-type-vulnerabilities:alpine:top20} Top 20 datasets.}\label{tab:package-type-vulnerabilities}
\end{figure*}

However, it cannot be ruled out that the same CVE applies to unrelated packages, namely, those with different upstreams. %
For this reason, we analyzed the types of packages with the same CVEs, reported in Figure~\ref{tab:package-type-vulnerabilities}.

The results clearly show that Docker, Google, Microsoft, and Amazon only report CVEs for upstream packages, while the other tools duplicate the same CVE for packages with the same upstream. Furthermore, tools correctly recognize cases where different upstreams are subjected to the same CVEs. This is more apparent for the datasets containing all packages, as the others share a certain amount of packages.

\begin{observation}
    Trivy and Anchore duplicate the same CVEs for every package with the same upstream. %
\end{observation}

Finally, Figure~\ref{fig:cve-vulnerabilities} characterizes the differences in the CVEs detected by the different tools. The figure shows that all tools produce a comparable number of vulnerabilities for Alpine excluding Anchore. In fact, Anchore reports about 50\% more CVEs than the others. Instead, all tools reports varying number of CVEs for Debian; Amazon and Microsoft report the fewest CVEs.

\begin{figure*}[tbp!]
    \centering
    \subcaptionbox{\label{fig:cve-vulnerabilities:debian:all}}{%
        \heatmap[-20]{%
            Amazon,Anchore,Docker,Gcloud,Microsoft,Trivy%
        }{%
            Amazon,Anchore,Docker,Gcloud,Microsoft,Trivy%
        }{%
            {1.00,0.19,0.07,0.06,0.01,0.19},%
            {0.19,1.00,0.34,0.72,0.01,0.86},%
            {0.07,0.34,1.00,0.47,0.01,0.40},%
            {0.06,0.72,0.47,1.00,0.00,0.81},%
            {0.01,0.01,0.01,0.00,1.00,0.00},%
            {0.19,0.86,0.40,0.81,0.00,1.00},%
        }%
    }%
    \hfill%
    \subcaptionbox{\label{fig:cve-vulnerabilities:alpine:all}}{%
        \heatmap[-20]{%
            Amazon,Anchore,Docker,Gcloud,Microsoft,Trivy%
        }{%
            Amazon,Anchore,Docker,Gcloud,Microsoft,Trivy%
        }{%
            {1.00,0.52,0.99,0.97,0.99,0.99},%
            {0.52,1.00,0.53,0.51,0.51,0.51},%
            {0.99,0.53,1.00,0.98,0.98,0.98},%
            {0.97,0.51,0.98,1.00,0.96,0.96},%
            {0.99,0.51,0.98,0.96,1.00,1.00},%
            {0.99,0.51,0.98,0.96,1.00,1.00},%
        }%
    }
    \hfill%
    \subcaptionbox{\label{fig:cve-vulnerabilities:debian:top20}}{%
        \heatmap[-20]{%
            Amazon,Anchore,Docker,Gcloud,Microsoft,Trivy%
        }{%
            Amazon,Anchore,Docker,Gcloud,Microsoft,Trivy%
        }{%
            {1.00,0.04,0.04,0.04,0.01,0.84},%
            {0.04,1.00,0.73,0.99,0.04,0.15},%
            {0.04,0.73,1.00,0.74,0.05,0.11},%
            {0.04,0.99,0.74,1.00,0.04,0.15},%
            {0.01,0.04,0.05,0.04,1.00,0.01},%
            {0.84,0.15,0.11,0.15,0.01,1.00},%
        }%
    }
    \hfill%
    \subcaptionbox{\label{fig:cve-vulnerabilities:alpine:top20}}{%
        \heatmap[-20]{%
            Amazon,Anchore,Docker,Gcloud,Microsoft,Trivy%
        }{%
            Amazon,Anchore,Docker,Gcloud,Microsoft,Trivy%
        }{%
            {1.00,0.19,0.82,0.64,1.00,1.00},%
            {0.19,1.00,0.15,0.14,0.19,0.19},%
            {0.82,0.15,1.00,0.78,0.82,0.82},%
            {0.64,0.14,0.78,1.00,0.64,0.64},%
            {1.00,0.19,0.82,0.64,1.00,1.00},%
            {1.00,0.19,0.82,0.64,1.00,1.00},%
        }%
    }
    \caption{Jaccard index for the different CVEs in the: Debian \subref{fig:cve-vulnerabilities:debian:all} All and \subref{fig:cve-vulnerabilities:debian:top20} Top 20 datasets; Alpine \subref{fig:cve-vulnerabilities:alpine:all} All and \subref{fig:cve-vulnerabilities:alpine:top20} Top 20 datasets.}\label{fig:cve-vulnerabilities}
\end{figure*}

\subsection{Common Vulnerabilities and Exposures}
\label{sec:accuracy:cves}

The number of reported CVEs %
significantly varies across tools, even by one order of magnitude (Table~\ref{tab:vulnerabilities}). %
Such discrepancies called for an analysis of how tools map packages to CVEs.

\subsubsection{Irrelevant CVEs}
\label{issue:cves:kernel}

The misuse of the upstream parameter led us to the hypothesis that tools may find irrelevant CVEs. One representative example is given by CVEs related to the Linux kernel. In fact, containers run on the host OS, so they are not affected by kernel-related issues. However, it is not uncommon for containers to include kernel headers since they might be needed by software components that interface with low-level features of the OS, such as system monitoring tools. We removed kernel packages from the SBOMs generated by the different tools and provided the related results in a separate column of Table~\ref{tab:pkgs}. The resulting number of CVEs significantly changed in all datasets. Docker, Google, and Microsoft show the fewest kernel-related CVEs across the four datasets, whereas all other tools report a large number of irrelevant CVEs.

\begin{observation}
    SCA tools for containers may incorrectly include irrelevant CVEs in the SBOMs, such as those that are kernel-related.
\end{observation}

\begin{table}
    \def\arraystretch{1.125}
    \setlength{\tabcolsep}{4.25pt}
    \centering\footnotesize\sffamily
    \codefontsize
    \subcaptionbox{\label{tab:cve:all}}{%
    \begin{tabular}{llrrrrrrr}
    \toprule
    & & \scalebox{.95}[1.0]{\textbf{True}} & \scalebox{.95}[1.0]{\textbf{False}} & \scalebox{.95}[1.0]{\textbf{False}} \\[-2pt]
    \textbf{Dataset} & \textbf{Tool} & \textbf{Pos.} & \textbf{Pos.} & \textbf{Neg.} & \textbf{Prec.} & \scalebox{.9}[1.0]{\textbf{Recall}} & \textbf{F1} \\
    \cmidrule(lr){1-1} \cmidrule(lr){2-2} \cmidrule(lr){3-5} \cmidrule(lr){6-8}
    \multirow{6}{1cm}{\raggedleft\textbf{Debian All}} 
    & Amazon    &              178 &                15 &              2323 &        0.92 &     0.07 &       0.13 \\
    & Anchore   &             2227 &                31 &               274 &        0.99 &     0.89 &       0.94 \\
    & Docker    &             1226 &                66 &              1275 &        0.95 &     0.49 &       0.65 \\
    & Gcloud    &             2426 &                42 &                75 &        0.98 &     0.97 &       0.98 \\
    & Microsoft &                6 &               197 &              2495 &        0.03 &     0.00 &       0.00 \\
    & Trivy     &             2432 &                41 &                69 &        0.98 &     0.97 &       0.98 \\
    \cmidrule(lr){1-1} \cmidrule(lr){2-2} \cmidrule(lr){3-5} \cmidrule(lr){6-8}
    \multirow{6}{1cm}{\raggedleft\textbf{Alpine All}} 
    & Amazon    &               87 &                11 &                26 &        0.89 &     0.77 &       0.82 \\
    & Anchore   &               86 &                71 &                26 &        0.55 &     0.77 &       0.64 \\
    & Docker    &               92 &                11 &                21 &        0.89 &     0.81 &       0.85 \\
    & Gcloud    &               88 &                11 &                25 &        0.89 &     0.78 &       0.83 \\
    & Microsoft &               87 &                11 &                26 &        0.89 &     0.77 &       0.82 \\
    & Trivy     &               88 &                11 &                25 &        0.89 &     0.78 &       0.83 \\
    \cmidrule(lr){1-1} \cmidrule(lr){2-2} \cmidrule(lr){3-5} \cmidrule(lr){6-8}
    \multirow{6}{1cm}{\raggedleft\textbf{Debian Top 20}}
    & Amazon    &               63 &                 1 &               198 &        0.98 &     0.24 &       0.39 \\
    & Anchore   &              261 &                 2 &                 0 &        0.99 &     1.00 &       1.00 \\
    & Docker    &              198 &                 9 &                63 &        0.96 &     0.76 &       0.85 \\
    & Gcloud    &              261 &                 2 &                 0 &        0.99 &     1.00 &       1.00 \\
    & Microsoft &               11 &                 0 &               250 &        1.00 &     0.04 &       0.08 \\
    & Trivy     &              261 &                 2 &                 0 &        0.99 &     1.00 &       1.00 \\
    \cmidrule(lr){1-1} \cmidrule(lr){2-2} \cmidrule(lr){3-5} \cmidrule(lr){6-8}
    \multirow{6}{1cm}{\raggedleft\textbf{Alpine Top 20}} 
    & Amazon    &               11 &                 0 &                67 &        1.00 &     0.14 &       0.25 \\
    & Anchore   &               19 &                32 &                59 &        0.37 &     0.24 &       0.29 \\
    & Docker    &                9 &                 0 &                69 &        1.00 &     0.12 &       0.21 \\
    & Gcloud    &                7 &                 0 &                71 &        1.00 &     0.09 &       0.16 \\
    & Microsoft &               11 &                 0 &                67 &        1.00 &     0.14 &       0.25 \\
    & Trivy     &               11 &                 0 &                67 &        1.00 &     0.14 &       0.25 \\
    \bottomrule
    \end{tabular}
    }
    \caption{Classification performance for the considered datasets.}\label{tab:classification-performance}
\end{table}

\subsubsection{Classification Performance}
\label{issue:cves:docker}

The individual CVEs reported by the different tools have the same order of magnitude in all datasets, excluding Microsoft and Amazon. We further analyzed these CVEs by deriving the key metrics characterizing classification performance, reported in Table~\ref{tab:classification-performance}. Among them, the number of false negatives is the most important because it means that the container has vulnerabilities that are \emph{not} detected, possibly leading to security risks.

We noticed a significant discrepancy between the number of vulnerabilities in Debian and Alpine. This can be explained by the fact that Debian offers approximately five times more packages than Alpine, which corresponds to a similar difference in the number of vulnerabilities. Additionally, Alpine releases a new version approximately every six months~\cite{AlPINE:2025:RELEASE} while Debian does that every two years~\cite{DEBIAN:2025:RELEASE}.

\paragraph{False Positives}

The analysis of false positives reveals different patterns in Alpine and Debian. For Alpine, we observe a relatively consistent number of vulnerabilities reported as false positives. These primarily arise because the tools rely on external advisories or methods (e.g., CPE analysis) that are not well aligned with Alpine's versioning schema. This is mostly noticeable for Anchore in the Alpine Top 20 dataset.

In contrast, the sources of false positives for Debian are more diverse. For Amazon and Microsoft, incorrect package identifiers lead to an inaccurate number of CVEs being reported. Specifically, Microsoft lacks upstream or distribution-level information, which is essential in the case of Debian, while Amazon reports incorrect upstreams. For the remaining tools, false positives generally stem from misinterpretations of the information provided in security trackers. Docker represents a special case, as it performs string matching from the notes in the security tracker that are associated with the CVE reports in Debian, resulting in substantial false positives.

\paragraph{False Negatives}

The analysis of false negatives also reveals distinct behaviors in Alpine and Debian. %
In Alpine, false negatives arise from an inaccurate selection of the source of truth. Specifically, tools often rely on a database~\cite{ALPINE:2025:SECDB} that lists packages only when a given CVE has been explicitly confirmed as effective. As a result, packages that are possibly vulnerable but not yet tested are omitted.

In Debian, false negatives consistently occur when the package version contains a mixture of numbers and strings (e.g., \texttt{2.9.14\%2Bdfsg-1.3\textasciitilde{}deb12u1}). This suggests that the issue is related to incorrect parsing and sorting of version identifiers. As with false positives, the results from Microsoft and Amazon are less accurate due to incorrect information extracted from the container. Similar to the previous case, Docker also exhibit a large amount of false negatives for Debian, marking packages as non-vulnerable when they are actually vulnerable but do not require an immediate security update.

%% file: vulnerability.tex
\section{SBOM Confusion Vulnerability}
\label{sec:vuln}

The differences in the content of the SBOMs across tools introduce critical incompatibilities, as identified in Section~\ref{sec:format}. These issues are particularly alarming because users should be able to rely on SBOMs to accurately %
characterize the contents of container images, especially those created by third parties.

Let us consider current practices in software development for cloud-native application. Using a container image (either as a base image or as it is) entails generating or retrieving the corresponding SBOM. Such a SBOM is typically deployed into an artifact registry and linked to a CI\,/CD pipeline or a security platform which periodically scan them to find and mitigate newly disclosed vulnerabilities. However, problems may arise when the SBOM was produced with one tool and is scanned by another tool. In fact,
inconsistencies in how these tools represent package identifiers result in missing or misrepresenting vulnerabilities. Consequently, the scanning may report significantly fewer vulnerabilities than those actually existing, if at all. We refer to this security issue as the \textit{SBOM confusion vulnerability}. Note that it is not the result of a malicious actor but rather a byproduct of the fragmented SSC ecosystem, where heterogeneous tooling produces incompatible SBOM outputs.

\begin{figure}[tbp!]
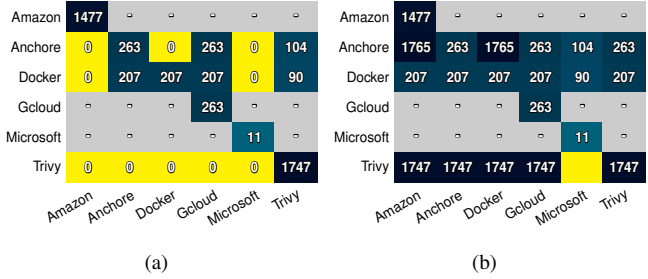

    \centering
    \hspace{-7.5pt}%
    \subcaptionbox{\label{fig:sbomvert:before}}{%
        \heatmap[-20]{%
            Amazon,Anchore,Docker,Gcloud,Microsoft,Trivy%
        }{%
            Amazon,Anchore,Docker,Gcloud,Microsoft,Trivy%
        }{%
            {1477,-,-,-,-,-},%
            {0,263,0,263,0,104},%
            {0,207,207,207,0,90},%
            {-,-,-,263,-,-},%
            {-,-,-,-,11,-},%
            {0,0,0,0,0,1747},%
        }%
    }%
    \hspace{0pt}%
    \subcaptionbox{\label{fig:sbomvert:after}}{%
        \heatmap[-20]{%
            Amazon,Anchore,Docker,Gcloud,Microsoft,Trivy%
        }{%
            Amazon,Anchore,Docker,Gcloud,Microsoft,Trivy%
        }{%
        {1477, -,   -,   -,   -,   -},%
        {1765, 263, 1765, 263, 104, 263},%
        {207,  207, 207, 207,  90, 207},%
        {-,    -,   -,   263,  -,   -},%
        {-,    -,   -,   -,   11,  -},%
        {1747,1747,1747,1747,,1747},%
        }%
    }%
    \caption{Unique CVEs found by the different tools on the Debian Top 20 dataset \subref{fig:sbomvert:before} before and \subref{fig:sbomvert:after} after running \texttt{sbomvert}.}\label{fig:sbomvert}
\end{figure}

To quantify the resulting impact, we measured the number of Common Vulnerabilities and Exposures (CVEs) detected when each tool parsed SBOMs generated by other tools. %
Figure~\ref{fig:sbomvert:before} shows the related results, which highlight that no vulnerabilities were detected across different tool combinations in many cases. Moreover, the number of detected vulnerabilities was always lower than when the SBOM was analyzed with the same or matching tool. These findings highlight that SBOM incompatibilities can significantly undermine %
security.

\subsection{Mitigation: \texttt{sbomvert}}

Based on the issues we have previously identified (summarized by Table~\ref{tab:issues} in appendix), we devised a set of rules to translate an SBOM produced by one tool into one compatible with another tool, thereby enabling correct vulnerability detection. Both the input and output SBOMs are in SPDX format, ensuring that no information is lost during the translation process. We implemented this mechanism as an open-source tool %
called \texttt{sbomvert}.

We evaluate its effectiveness by re-deriving the number of unique CVEs reported by the different tools, this time after translating the source SBOM with \texttt{sbomvert} into one suitable for the specific vulnerability scanner. The obtained results are shown in Figure~\ref{fig:sbomvert:after}, which demonstrates a substantial improvement in the number of detected vulnerabilities. In many cases, the number of detected CVEs across tools aligns closely with that of the original SBOM. Clearly, the effectiveness of the translation depends on the completeness of the source SBOM. For instance, SBOMs generated by Microsoft's \texttt{sbom-tool} omit information about the OS distribution, which reduces the number of vulnerabilities that can be detected even after translation. Nevertheless, \texttt{sbomvert} provides a practical mechanism to mitigate the inconsistencies due to heterogeneous tooling and improve the reliability of SBOM-based vulnerability monitoring.

%% file: discussion.tex
\section{Discussion}
\label{sec:discussion}

This section discusses the limitations of our work as well as additional findings. Suggestions on how to improve security trackers are further provided in Appendix~\ref{sec:security-trackers}.

\paragraph{Limitations and Threat to Validity}

We restrict our attention to SBOMs in the SPDX format because it is supported by all the tools we considered. Then, we only focus on Debian and Alpine images because they are widespread and they are the de facto standard in many of the top downloaded container images. The dataset with all the Debian packages does not contain many vulnerable packages because it was built in 2025, and we decided to remove all CVEs discovered in 2025. Therefore, the results on F1-score, precision, and recall may not be entirely representative of the actual CVE detection performance of the tools. However, they are still valid in showing that both false positives and false negatives occur across tools. Moreover, we excluded vulnerabilities without a CVE ID, which may result in a different output than the one produced by the tools in their default configuration.

\paragraph{SBOM Translation}

We were able to improve the overall interoperability between SBOM formats with \texttt{sbomvert}. However, Figure~\ref{fig:sbomvert} clearly shows that a few challenges still remain. In fact, the mapping between packages and CVEs becomes lossy (e.g., in the case of Microsoft's \texttt{sbom-tool}) when pURLs are incomplete. Conversely, the number of reported CVEs increases when the SBOM generator includes both binary and source packages. Overall, SBOM generators should report only the packages actually installed %
and also include the necessary information to correctly identify them.

\paragraph{CVE Formats}

The number of reported CVEs depends not only on the actual %
vulnerabilities affecting a given container but also on how the CVEs are presented. Table~\ref{tab:pkgs} shows that, in many cases, the total number of reported CVEs and the number of distinct CVEs vary significantly across tools. In fact, CVEs may be reported either per binary package or per source package (see Table~\ref{tab:package-type-vulnerabilities}). %
The latter is particularly problematic, as not all binary packages built from a vulnerable source are necessarily affected by a given CVE. This can lead to inflated or misleading vulnerability counts. Overall, the discrepancy between CVE formats creates confusion, as it becomes difficult to compare results across tools and accurately assess the security posture of a container.

\paragraph{Zero-CVE Container Images}

A recent trend in the container security industry is the use of zero-CVE images, which are images with no known CVEs. This is achieved by reducing the number of installed packages to a minimum and patching CVEs that have not yet been merged in the upstream packages. We are convinced that this approach can greatly reduce the attack surface of container images; however, it is still prone to SBOM inconsistencies. Additionally, it is unclear how such images identify CVEs (e.g., by relying on advisories) and which package versions and CVEs have been manually patched. %
Adding the actual patches and vulnerability advisories %
would significantly contribute to the transparency and trust in zero-CVE images. This would also facilitate reproducibility and %
allow for more accurate vulnerability scanning. %

%% file: related.tex
\section{Related Work}
\label{sec:rw}
\label{sec:related}

There is a large share of works on supply chain security~\cite{Williams:2025:TESM}, including on threats~\cite{Hammi:2023:CSUR},  attacks~\cite{LADISA:2023:IEEESP}, and SBOM-related issues~\cite{Bi:2024:TSEM}. We summarize those that are more relevant next.

Duan et al. \cite{DUAN:2021:NDSS} introduce a %
framework to assess the functional and security features of package managers for interpreted languages. Their solution leverages program analysis techniques to study registry abuse related to different types of supply chain attacks. For this reason, their approach relies on the availability of the application source code. Instead, we target binary OS packages %
installed in container images.

Yu et al.~\cite{Yu:24:DSN} focus on the correctness of SBOM generation, defined in terms of discrepancies between the number of reported packages, their similarity, and the presence of duplicates. %
However, they do not cover scanning and primarily consider programming language packages in their analysis, as opposed to the OS packages we target in this work.

Rabbi et al.~\cite{RABBI:2024:SIGAPP} conduct a detailed evaluation of SBOM generation tools for \texttt{npm} packages, including accuracy and precision\,/\,recall. They find that several tools are unable to correctly detect dependencies, posing a definite threat for real-world applications. However, they do not analyze the reasons behind these inconsistencies as we do in this work.

Xia et al.~\cite{Xia:2023:ICSE} carry out an empirical study on the developer perceptions of SBOMs. Specifically, they gathered data on both SBOM-related practices, tooling, and concerns through both interviews and an online survey, overall involving 82 practitioners. Their findings highlighted that SBOM tooling is still immature, calling for high-quality, standard-compliant, and interoperable tools. %
Stalnaker et al.~\cite{Stalnaker:2024:ICSE} report similar observations from another study. Unfortunately, they do not consider specific tools suitable for container images, nor provide a quantitative evaluation of their accuracy and compliance.

Halbritter and Merli~\cite{Halbritter24} evaluate the accuracy and reliability of SBOM tools, with a special focus on the compliance with the National Telecommunications and Information Administration (NTIA) requirements. However, they specifically address web applications (written in Python or Typescript) and system software (in C or Rust) instead of container images. Torres-Arias et al.~\cite{Torres-Arias:2023:SP} provide a similar analysis based on the public \texttt{bom-shelter} dataset, which includes SBOM generated for containers. Again, their results are primarily concerned about NTIA compliance.

Kim et al.~\cite{kimWhyJohnnyCan2023} carry out a usability study of different tools to scan vulnerabilities of container images. As a result, they %
find that the considered tools provide information that are ambiguous, incomplete, or difficult to act upon. Their findings are based on heuristics and mainly qualitative. Instead, this work takes a quantitative approach to understand the effectiveness of SCA tools, including their accuracy.

Kawaguchi et al.~\cite{Kawaguchi24} target vulnerability analysis of packages that are manually installed in container images, namely, without using a package manager. A comprehensive data-driven evaluation revealed that scanners fail to identify up to 70\% of these packages, resulting in vulnerabilities that are largely undetected. In contrast, we show that employing different tools for SBOM generation and vulnerability analysis leads to a significant variation in the number and type of security issues that can be found for OS packages.

O'Donoghue et al.~\cite{ODonoghue:2024:SCORED} investigate the impact of using specific combinations of tools to generate SBOMs for containers and scan the corresponding vulnerabilities. They found a high variability in the number of reported vulnerabilities, %
however, they did not carry out a detailed analysis of the causes behind such a variability as we instead do here. We also provide \texttt{sbomvert} as a simple solution to enable interoperability in practice. %

Dalia et al.~\cite{Dalia:2024:ARES} compare different tools for SBOM generation, primarily in terms of features (e.g., the support for different platforms and their integration with) and qualitative metrics (e.g., user friendliness). Unfortunately, their analysis does not include a data-driven evaluation, neither considers the impact of SBOM generation on vulnerability assessment.

%% file: conclusion.tex
\section{Conclusion}
\label{sec:conclusion}

This paper addressed the issue of SBOM incompatibility and discussed the reasons why different SCA tools for containers report varying operating system packages and vulnerabilities. We began by demonstrating that SCA tools are not interoperable and using different tools for creating SBOMs and for scanning them leads to inaccurate results. We then analyzed the differences in SBOM formats across various tools. We employed this information to generate a common package index, which helped us examine the discrepancies in package and vulnerability detection. Our evaluation indicated that the tools largely identify the same packages, and the distinct number of discovered CVEs is often of the same order of magnitude. Additionally, we demonstrated that the tools may inflate the number of discovered CVEs by including irrelevant packages. Finally, we employed the results from the package difference analysis to develop a tool, \texttt{sbomvert}, which translates SBOMs and pURLs to enable interoperability between tools. We hope that our work can serve as a foundation for more standardized %
formats and motivate companies to improve their tooling.

%% file: appendix.tex
\section{SPDX Package in Trivy}

Listing~\ref{fig:trivy} shows an SPDX package generated by Trivy for the \texttt{bsdutils} software as distributed by Debian. We notice that package information is shown in three distinct locations: \texttt{sourceInfo}, \texttt{attributionTexts} and \texttt{referenceLocator}. We found that Trivy ignores the pURL and the other standard fields in the package for vulnerability detection and relies on storing package information in the optional \texttt{sourceInfo} field with the format:
\begin{center}
    \observationfontsize
    \texttt{built package from: <package-name> <package-version>}
\end{center}
Note that this non-standard parameter has indeed a different content than the version in the pURL. In fact, the package version of the \texttt{sourceInfo} optional fields has to be in the form \texttt{epoch:version} and the package name should be set to the Debian upstream that corresponds to the Debian source package (\texttt{util-linux} in this example).\vspace{.25em}

\begin{figure}[htb!]
\begin{lstlisting}[style=spdx,moreattributes={name, SPDXID, versionInfo, sourceInfo, externalRefs, referenceCategory, referenceType, referenceLocator, attributionTexts, primaryPackagePurpose}]
"(*@\color{aaltoGreen}name@*)": "bsdutils",
"SPDXID": "SPDXRef-Package-c08b550a1bd747d6",
"versionInfo": "1:2.38.1-5+deb12u3",
"sourceInfo": 
"built package from: util-linux 2.38.1-5+deb12u3",
"externalRefs": [{
  "referenceCategory": "PACKAGE-MANAGER",
  "referenceType": "purl",
  "referenceLocator": 
  "pkg:deb/debian/bsdutils@2.38.1-5%2Bdeb12u3?
  arch=amd64\u0026distro=debian-12.10\u0026epoch=1"
}
],
"attributionTexts": [
  "LayerDiffID: sha256:..",
  "PkgID: bsdutils@1:2.38.1-5+deb12u3",
  "PkgType: debian"
],
"primaryPackagePurpose": "LIBRARY"
\end{lstlisting}%
  \caption{Sample SPDX Package content in Trivy.}\label{fig:trivy}
\end{figure}

\section{Comparison between Debian and Alpine}
\label{sec:comparison}

We compared the detected packages and CVEs across the Top 20 containers in Debian and Alpine. First, we notice that the number of packages in each container is approximately five times larger than the Alpine counterpart. We noticed the same pattern for the number of CVEs (see Table~\ref{tab:pkgs}). We also observed that in most cases, the package versions of Alpine were more up-to-date and therefore with fewer vulnerabilities. This is due to the faster Alpine release cycle.

We then checked if the security advisories of the different OSes were showing the same vulnerabilities. For that, we employed the Debian Security Tracker database and looked at similar versions of the same package %
for Debian and Alpine. %
We observed that the Alpine Security Tracker (AST) occasionally omits information on vulnerable packages. For example, \texttt{wget@1.24.5-r0} for Alpine~3.20 is listed with no known CVEs in the AST, while the Debian Security Tracker for Bookworm (Debian 12) reports it as vulnerable to \texttt{CVE-2024-10524} and \texttt{CVE-2021-31879}. This discrepancy likely reflects a decision not to issue hotfix updates for these CVEs in older package versions.

Our analysis highlights a key architectural difference between the two systems. The Debian Security Advisory (DSA) tracks all CVEs affecting packages in supported releases, whereas the AST only records vulnerabilities once they have been fixed at the individual release level. This explains why these CVEs are absent from the 3.20 tracker but are present in the 3.21 tracker.

\section{Improvements to %
Security Trackers}
\label{sec:security-trackers}

We found many opportunities to improve the Debian Security Tracker (DST) and Alpine Security Tracker (AST), which we detail next.

\paragraph*{Missing Information}

Both DST and AST track vulnerabilities only at the source package level. In other words, if a user installs only a subset of binaries from a source \texttt{dpkg} package that has known vulnerabilities, it is impossible to determine whether those specific binaries are affected or not. %

Furthermore, trackers do not include machine-readable information about the processor architecture vulnerable to a specific CVE. This leaves room for many false positives.

\paragraph*{Fine-grained Description of CVE Applicability}

A CVE identifier alone is not sufficient to determine whether a vulnerability applies to a specific use case. For instance, a bug that crashes a CLI tool in an operating system environment could lead to a denial-of-service (DoS) attack in a microservice using a container with that tool. 

The DST assigns the ``unimportant'' label to a broad set of CVEs that are not considered applicable or non-immediately exploitable. This means that packages with CVEs targeting a different operating system or that require a long list of incorrect parameters to be set or that require the user to be root will all be treated in the same way. The AST does not provide any note or textual information of the reasons a CVE applies to a package. This complicates CVE triaging and planning. Emerging approaches such as VEX~\footnotetext{The Vulnerability Exploitability eXchange (VEX) is a standard format used to add context information about a CVE by assigning a status (e.g., non-affected) and a justification (e.g., compilation flag not used in our build).}statements~\cite{NTIA:2021:VEX} clarify the applicability of CVEs and would enable security trackers to convey contextual information that supports more accurate risk prioritization and mitigation.

\paragraph{Missing Endpoints to Establish Ground Truth}

Debian does not track all deployed versions of each package in its security database. Instead, it records only the versions that are known to be vulnerable or that include fixes for vulnerabilities. This makes determining whether a specific package version is vulnerable less straightforward. We believe that the entire ecosystem would benefit from an endpoint that can indicate whether a given package version has any known vulnerabilities.

Instead, Alpine has a specific endpoint to track vulnerable packages. However, old versions do not seem to be included (as shown in Appendix~\ref{sec:comparison}) leading to the false negatives reported by many tools.

\begin{table*}[!htb]
    \centering
    \def\arraystretch{1.125}
    \setlength{\tabcolsep}{4.25pt}
    \centering\footnotesize\sffamily
    \codefontsize
    \begin{tabular}{llll}
        \toprule
        \textbf{Source} & \textbf{Issue} & \textbf{Affected Tools} & \textbf{Section} \\
        \midrule
        \multirow{3}{*}{pURL} 
         & pURL is invalid                & All       & \ref{issue:purl:invalid} \\
         & pURL is incomplete             & Microsoft & \ref{issue:purl:incomplete} \\
         & pURL has incorrect information & Amazon    & \ref{issue:purl:incorrect} \\
        \midrule
        \multirow{2}{*}{SBOM}
         & pURL not used as package identifier & Trivy, Anchore          & \ref{issue:format:nopurl} \\
         & Reliance on optional fields or custom entries & Docker, Trivy & \ref{issue:format:fields} \\
        \midrule
        \multirow{4}{*}{CVEs}
         & Source Binary Mismatch & All                            & \ref{sec:source-binary-mismatch} \\
         & Irrelevant CVEs & Trivy, Amazon, Microsoft & \ref{issue:cves:kernel} \\
         & Incorrect reporting of CVEs & Docker                    & \ref{issue:cves:docker} \\ 
        \bottomrule
    \end{tabular}
    \caption{Summary of issues and affected tools.}\label{tab:issues}
\end{table*}